\documentclass[10pt,letterpaper,twocolumn]{article} 

\usepackage{ol2}
\usepackage[draft]{hyperref}
\usepackage{amsmath}

\begin{document}

\twocolumn[ 

\title{Dynamical trapping of light in modulated waveguide lattices}


\author{Stefano Longhi}

\address{Dipartimento di Fisica, Politecnico di Milano, Piazza L. da Vinci 32, I-20133 Milano, Italy}

\begin{abstract}
A discrete analogue of the
 dynamical (Kapitza) trapping effect, known for classical
and quantum particles in rapidly oscillating potentials, is proposed
for light waves in modulated graded-index waveguide lattices. As in
the non-modulated waveguide lattice a graded-index potential can
confine light at either normal or Bragg angle incidence, periodic
modulation of the potential in the longitudinal direction enables to
trap optical beams at both normal and Bragg incidence angles.
\end{abstract}



 ] 

\noindent Light propagation in photonic lattices has attracted a
great interest over the past few years, with the observation of a
host of new phenomena with no counterpart in continuous media
\cite{review1,review2}. Flexible control of light transport and
localization in such devices can be realized by breaking the
translation invariance of the lattice along the propagation
direction by either periodic axis bending or out-of-phase modulation
of refractive index of adjacent guides. Examples of light control in
modulated lattices include dynamic localization
\cite{Longhi05,Longhi06,Iyer07,Szameit10}, tunneling inhibition
\cite{Szameit09,Kartashov09,Kartashov10}, multiband refraction
control \cite{Longhi06OL}, polychromatic diffraction management
\cite{Garanovich06,SzameitNature09,Garanovich10}, and defect-free
surface waves \cite{Garanovich08,Szameit08}. Most of such light
control techniques bear interesting analogies with coherent control
of driven quantum systems, such as electronic or matter wave
transport in driven lattices \cite{Longhi09LPR}.\\
In this Letter a mechanism of light trapping in graded-index
modulated waveguide lattices is proposed, which is based on a {\em
discrete analogue} of the Kapitza (or dynamical) stabilization
effect of classical and quantum particles in rapidly oscillating
potentials \cite{K1,K2,K3,K4,K5}. Dynamical stabilization generally
refers to the possibility for a particle to be trapped by a
rapidly-oscillating potential in cases where the static potential
cannot. Well-known paradigms are the Kapitza stabilization of the
pendulum \cite{K1,K2} and Paul traps for charged particles
\cite{K3}. Here it is shown that modulated lattices can trap light
beams under conditions where the non-modulated lattice cannot. Let
us consider light propagation in a modulated waveguide lattice
\cite{Szameit09}, which in the tight-binding approximation is
governed by the discrete Schr\"{o}dinger equation
\begin{equation}
i \frac{d \psi_n}{dz}=-\kappa( \psi_{n+1}+\psi_{n-1})+\Phi_n(z)
\psi_n \; ,
\end{equation}
where $\psi_n(z)$ is the light field amplitude trapped in the $n$th
waveguide, $z$ is the longitudinal propagation coordinate,
$\kappa>0$ is the coupling constant between adjacent waveguides, and
$\Phi_n(z)$ is the longitudinal modulation of the propagation
constant for the $n$th guide. In the following, we will assume
$\Phi_n(z)=f(z)V_n$, where $f(z)$ is a modulation function with
spatial frequency $\omega=2 \pi / \Lambda$ and zero mean, and $V_n$
is the graded-index (static) potential. In waveguide arrays
manufactured by femtosecond laser writing, the modulation can be
realized by slightly varying the writing speed for each waveguide
\cite{Szameit09}. In the absence of the longitudinal modulation,
i.e. for $f=1$, a graded-index potential $V_n$ can be exploited to
focus or trap light beams. Such a kind of focusing has been
proposed, for example, to achieve deep subwavelength focusing in
metal-dielectric waveguide arrays \cite{Fan09}. A schematic of the
refractive index profile for a non-modulated graded-index array is
depicted in Fig.1. To relate the propagation properties of the
graded-index modulated lattice (1) with the dynamics of a quantum
particle in a rapidly-oscillating potential \cite{K4,K5}, it is
worth observing that the solution to the coupled-mode equations (1)
can be written as $\psi_n(z)=\psi(x=n,z)$, where the continuous
function $\psi(x,z)$ satisfies the Schr\"{o}dinger equation $i
(\partial \psi / \partial z)=\mathcal{H} \psi$ with Hamiltonian
\cite{Longhi07}
\begin{equation}
\mathcal{H}=-2 \kappa \cos (p_x)+ f(z) V(x)
\end{equation}
where $p_x=-i \partial/ \partial_x$ and $V(x=n)=V_n$.
 For a potential $V$ that varies slowly over one lattice period and for a
 wave packet with a narrow angular spectrum distribution, the semiclassical (ray
optics) equations for the mean values $\langle x \rangle $ of wave
packet position (in units of the array period) and $\langle
p_x\rangle$ of refraction angle, as given by the Ehrenfest theorem,
read \cite{Longhi07}
\begin{equation}
\frac{d \langle x \rangle}{dz} \simeq 2 \kappa \sin (\langle p_x
\rangle) \; \; , \;  \frac{d \langle p_ x \rangle  }{dz} \simeq
-f(z) \left( \frac{\partial V}{
\partial x} \right) (\langle x \rangle,z).
\end{equation}
\begin{figure}[htb]
\centerline{\includegraphics[width=8.2cm]{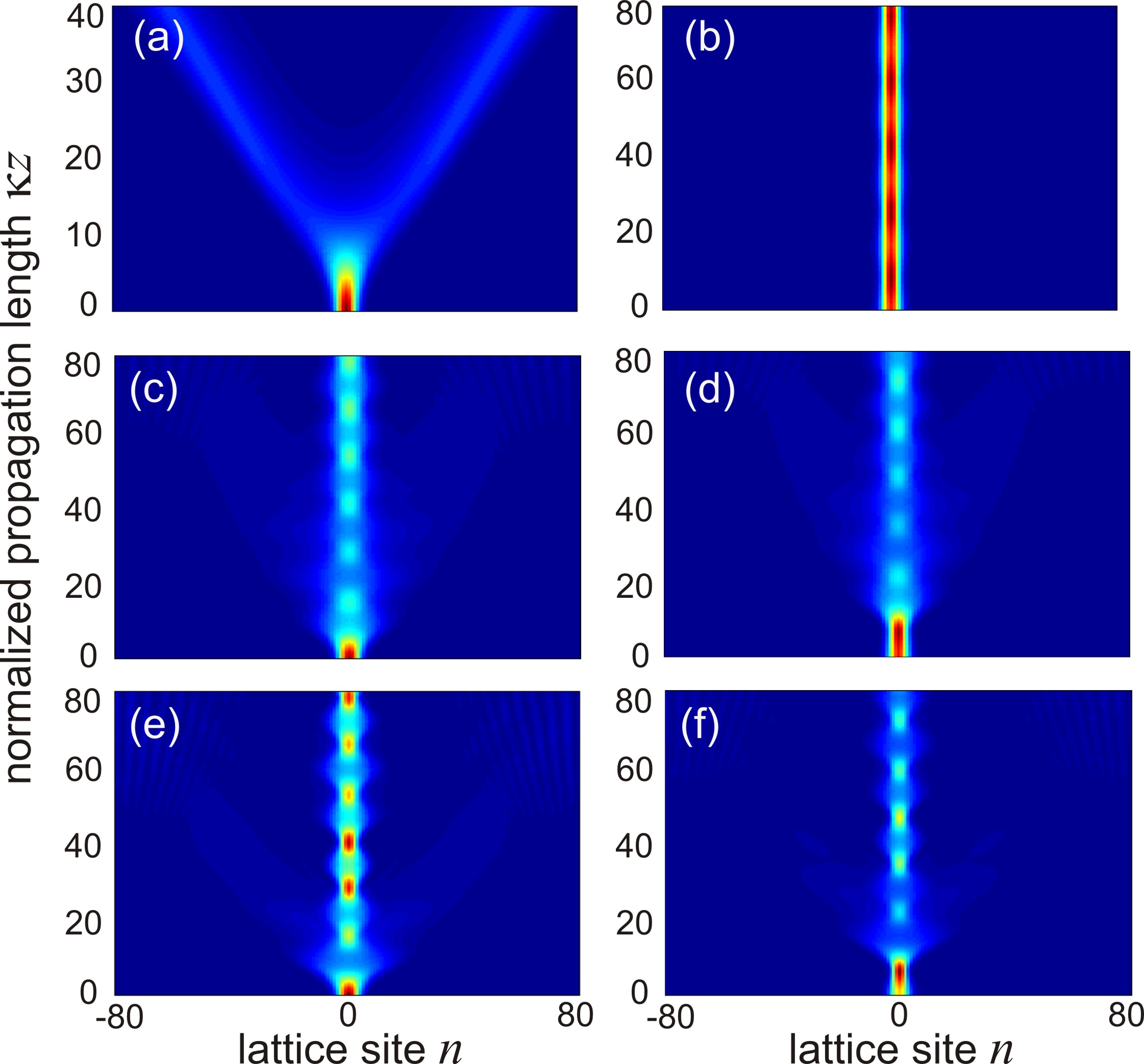}} \caption{ (Color
online) Beam propagation (snapshot of $|\psi_n(z)|^2$) in a
modulated lattice (square-wave modulation, $\omega / \kappa=0.5$)
with a Gaussian-shaped graded index potential (peak amplitude $V_0$)
for normal and Bragg incidence angles and for increasing values of
normalized potential peak amplitude: $V_0/ \kappa=0$ in (a) and (b)
(non-modulated lattice); $V_0 / \kappa=3$ in (c) and (d); $V_0 /
\kappa=5$ in (e) and (f). The right panels schematically show the
index profile of the non-modulated (upper plot) and modulated (lower
plot) graded-index waveguide arrays.}
\end{figure}
Let us assume that the static potential $V(x)$ is a bell-shaped
potential, for example, described by a parabolic or a Gaussian
function, with a maximum at $x=0$, and let us first consider the
non-modulated array ($f=1$). Then, according to Eqs.(3) there are
two fixed points, $(\langle x \rangle =0, \langle p_x \rangle =0)$
and $(\langle x \rangle =0, \langle p_x \rangle =\pi)$, the former
being unstable and the latter being stable \cite{note}. This means
that a broad beam launched into the array at nearly normal incidence
will not be trapped by the the static potential $V_n$, which acts as
a defocusing lens for the discretized beam. Conversely, an injected
broad beam tilted near the Bragg angle (corresponding to $\langle
p_x \rangle \simeq \pi$) will be confined by the potential $V_n$,
which acts as a focusing lens. This is shown, as an example, in
Figs.1(a) and (b). If the sign of the potential $V_n$ is reversed,
the stability of the two fixed points is interchanged. Such a
behavior is related to the well-known reversal of diffraction sign
for discretized light at normal or Bragg beam incidence
\cite{review1,Lederer02}. In fact, in the former case (normal
incidence) the Hamiltonian $\mathcal{H}$ is approximated as
$\mathcal{H} \simeq -\kappa
\partial^2_x- 2\kappa+V(x)$, whereas in the latter
case (Bragg angle incidence) one has $\mathcal{H} \simeq \kappa
\partial^2_x+ 2\kappa+V(x)$. Let us now consider the modulated
lattice, and show that, similarly to the dynamical trapping of
classical or quantum particles in rapidly-oscillating potentials
\cite{K1,K2,K4,K5}, dynamical beam  trapping in the lattice can be
realized for {\em both} normal and Bragg beam incidence. Indeed,
propagation of a broad beam, at either normal or Bragg incidence
angles, is governed by the Schr\"{o}dinger-type equation with an
oscillating potential, namely
\begin{equation}
i \frac{\partial \psi}{\partial z} \simeq \mp \kappa
\frac{\partial^2 \psi}{\partial x^2} \mp 2 \kappa \psi +f(z)V(x)
\psi
\end{equation}
where the upper (lower) sign applies to normal (Bragg) incidence.
For a rapidly oscillating potential, after setting
$\psi(x,z)=\phi(x,z) \exp \left[-i V(x) \left( \int_0^z d \xi
f(\xi)-\overline{\int_0^z d \xi f(\xi)} \right) \right]$ and
applying standard averaging methods \cite{K4,K5}, at leading order
the evolution of the slowly-varying envelope $\phi(x,z)$ is
described by a Schr\"{o}dinger-type equation with an effective
static potential $V_e(x)$, namely $i \partial_z \phi= [\mp \kappa
\partial^2_x \phi+V_e(x)]\phi$, where
\begin{equation}
V_e(x)=\mp 2 \kappa \pm \kappa \left( \frac{\partial V}{\partial
x}\right)^2 \overline{\left( \int_0^z d \xi f(\xi) -
\overline{\int_0^z d\xi f(\xi) }\right)^2}
\end{equation}
and the overbar denotes a spatial average over the oscillation
cycle. For a bell-shaped potential $V$ and considering normal beam
incidence, the effective potential $V_e(x)$ comprises two potential
barriers, which can support metastable (resonance) states (see, for
instance, \cite{K5}). Hence, though the static potential (i.e. for
$f=1$) can not trap light beams at normal incidence near $x=0$, in
the modulated lattice this is possible owing to the dynamical
(Kapitza) stabilization effect. For beam incidence at the Bragg
angle, the sign of both diffraction and effective potential are
reversed, and thus trapping is possible as well. It should be noted
that the strength of the confining part of the effective potential
is usually very small (it scales as $\sim 1/ \omega^2$
\cite{K4,K5}), and the observation of such a dynamical stabilization
effect for broad light beams in waveguide lattices might require
extremely long propagation distances, which are not accessible with
current waveguide array set-ups. Remarkably, we found that the
dynamical trapping effect persists even for relatively small spatial
modulation frequencies $\omega$, of the order or smaller than the
coupling constant $\kappa$, making dynamical trapping observable
with current waveguide array set-ups. As an example, Fig.1 shows
dynamical trapping of a broad Gaussian beam, at both normal and
Bragg incidence angles, for a Gaussian-shaped potential $V_n=V_0
\exp[-(n/w)^2]$ and for a square-wave modulation function [$f(z)=1$
in one semicycle, and $f(z)=-1$ in the other semicycle], as obtained
by direct numerical simulations of Eqs.(1) for $\omega/\kappa=0.5$,
$w=18$ and for increasing values of $V_0/\kappa$. Initial condition
is $c_n(0)= \exp(-n^2/25)$ in (a), (c) and (e) (normal incidence),
and $c_n(0)= (-1)^n \exp(-n^2/25)$ in (b),(d) and (f) (Bragg
incidence angle). For a typical coupling constant of $\kappa \simeq
0.5 \; {\rm mm}^{-1}$ \cite{Szameit09}, a propagation length of 40
in Fig.1 corresponds to a physical length of $\simeq 8$ cm.
Dynamical stabilization, observed in Figs.1(c-f), is related to the
existence of metastable (resonance) states of the modulated lattice,
which can be revealed by a direct computation of the quasienergy
spectrum and Floquet eigenstates of the coupled-mode equations (1)
with periodic coefficients. The quasi-energies (Floquet exponents)
$\mu$ and corresponding Floquet eigenstates $\psi_n^{(\mu)}(z)$ are
defined as the solutions to Eqs.(1) of the form
$\psi_n^{(\mu)}(z)=u_n^{(\mu)}(z) \exp(-i \mu z)$ with
$u_n^{(\mu)}(z+\Lambda)=u_n^{(\mu)}(z)$ and $ -\omega/2 \leq \mu <
\omega/2$. In a truncated lattice, a metastable (resonance) state to
Eqs.(1) corresponds to a Floquet state which is strongly localized
near $n=0$. For a modulation function $f(z)$ satisfying the symmetry
condition $f(-z)=f(z)$, as the one used in the simulations of Fig.1,
it can be readily shown that, if $\psi_n^{(\mu)}(z)=u_n^{(\mu)}(z)
\exp(-i \mu z)$ is a Floquet eigenstate with quasienergy $\mu$, then
$\psi_n^{(-\mu)}(z)=(-1)^n u_n^{(\mu)}(-z) \exp(i \mu z)$ is also a
Floquet eigenstate with quasienergy $-\mu$. Therefore, resonance
states appear in pairs. For the modulated lattice with parameter
values used in the simulations of Fig.1, a direct computation of the
Floquet exponents (assuming a truncated waveguide array comprising
$2N+1=161$ waveguides, from $n=-N$ to $n=N$) shows the existence of
one pair of metastable states with quasienergies $\pm 0.0284 \kappa$
at $V_0 / \kappa=3$, and of two pairs of metastable states
(quasienergies  $\simeq \pm 0.0533 \kappa$ and $\simeq \pm 0.1424
\kappa$) at $V_0 / \kappa=5$. As a general rule, for a fixed
modulation frequency the number of resonance states increases as the
amplitude $V_0$ of the static potential is increased, which is in
agreement with the prediction based on the cycle-averaged model: in
that case the number of resonances sustained by the double-barrier
effective static potential (5) increases as $V_0$ is increased. The
profiles of the two resonance states at $V_0 / \kappa=3$,
corresponding to the dynamical trapping of Figs.1(c) and (d), are
depicted in Fig.2. In this case, the metastable state
$\psi_n^{(-\mu)}$ with negative quasienergy is mainly excited by a
normal incidence input beam, whereas the metastable state
$\psi_n^{(\mu)}$ with positive
quasienergy is the mainly excited state when a beam incident at the Bragg angle is considered.\\

\begin{figure}[htb]
\centerline{\includegraphics[width=8.2cm]{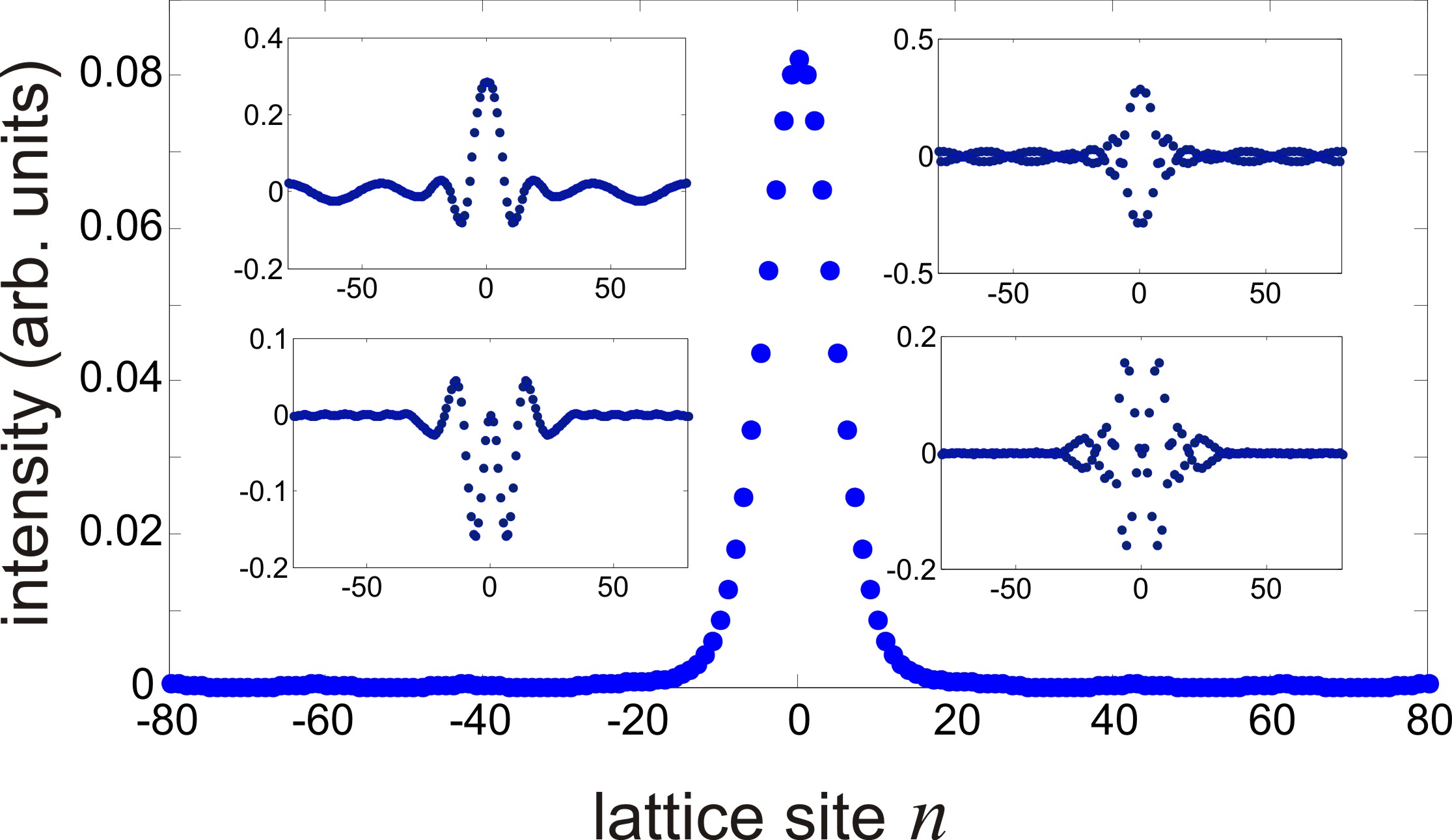}} \caption{ (Color
online) Numerically-computed intensity distribution $|\psi_n^{(\pm
\mu)}(z)|^2$ of the two metastable Floquet states, at the plane
$z=0$, for the modulated waveguide lattice of Fig.1 with $V_0/
\kappa=3$. The insets in the figure show the detailed behavior of
the real (upper panels) and imaginary (lower panels) parts of the
two Floquet states with negative (left plots) and positive (right
plots) quasienergy.}
\end{figure}

In conclusion, a discrete analogue of
 dynamical trapping of classical
or quantum particles in rapidly oscillating potentials has been
proposed for light waves in photonic lattices. Such an analogy could
be exploited to trap light beams at different incidence angles, thus
increasing the coupling efficiency from a broad angular spectrum
light source. However, as compared to ordinary focusing and trapping
in non-modulated graded-index lattices, dynamical trapping does not
sustain truly guided modes, and requires a more complex refractive
index management. Possible applications could be envisaged, for
example, in light trapping at the sub-wavelength regime \cite{Fan09}. \\
Work supported by the italian MIUR (Grant No. PRIN-2008-YCAAK).


\begin{thebibliography}{99}

\bibitem{review1}
F. Lederer, G. I. Stegeman, D. N. Christodoulides, G. Assanto, M.
Segev, and Y. Silberberg, Phys. Rep. {\bf 463}, 1 (2008).

\bibitem{review2}
 Y. V. Kartashov, V. A. Vysloukh, and L. Torner, Prog. Opt. {\bf 52}, 63 (2009).

\bibitem{Longhi05}
S. Longhi, Opt. Lett. {\bf 30}, 2137 (2005).

\bibitem{Longhi06}
S. Longhi, M. Marangoni, M. Lobino, R. Ramponi, P. Laporta, E.
Cianci, and V. Foglietti, Phys. Rev. Lett. {\bf 96}, 243901 (2006).

\bibitem{Iyer07}
R. Iyer, J.S. Aitchison, J. Wan, M.M. Dignam, and C.M de Sterke,
 Opt. Express {\bf 15}, 3212 (2007).

\bibitem{Szameit10}
A. Szameit, I.L. Garanovich, M. Heinrich, A.A. Sukhorukov, F.
Dreisow, T. Pertsch, S. Nolte,  A. T\"{u}nnermann, S. Longhi, and
Y.S. Kivshar, Phys. Rev. Lett. {\bf 104}, 223903 (2010).

\bibitem{Szameit09}
A. Szameit, Y.V. Kartashov, F. Dreisow, M. Heinrich, T. Pertsch, S.
Nolte, A. T\"{u}nnermann, V.A. Vysloukh, F. Lederer, and L. Torner,
 Phys. Rev. Lett. {\bf 102}, 153901 (2009).

\bibitem{Kartashov09}
Y.V. Kartashov, A. Szameit, V.A. Vysloukh, and L. Torner, Opt. Lett.
{\bf 34}, 2906 (2009).

\bibitem{Kartashov10}
Y.V. Kartashov and V.A. Vysloukh, Opt. Lett. {\bf 35}, 205 (2010).

\bibitem{Longhi06OL}
S. Longhi, Opt. Lett. {\bf 31}, 1857 (2006).

\bibitem{Garanovich06}
I.L. Garanovich, A.A. Sukhorukov, and Y.S. Kivshar, Phys. Rev. E
{\bf 74} 066609 (2006).

\bibitem{SzameitNature09}
A. Szameit, I.L. Garanovich, M. Heinrich, A.A. Sukhorukov, F.
Dreisow, T. Pertsch, S. Nolte, A. T\"{u}nnermann, and Y.S. Kivshar,
Nat. Phys. {\bf 5}, 271 (2009).

\bibitem{Garanovich10}
X.Y. Qi, I.L. Garanovich, A.A. Sukhorukov, W. Krolikowski, A.
Mitchell, G.Q. Zhang, D.N. Neshev, and Y.S. Kivshar, Opt. Lett. {\bf
35}, 1371 (2010).

\bibitem{Garanovich08}
I.L. Garanovich, A.A. Sukhorukov, and Y.S. Kivshar, Phys. Rev. Lett.
{\bf 100}, 203904 (2008).

\bibitem{Szameit08}
A. Szameit, I.L. Garanovich, M. Heinrich, A.A. Sukhorukov, F.
Dreisow, T. Pertsch, S. Nolte, A. T\"{u}nnermann, and Y.S. Kivshar,
Phys. Rev. Lett. {\bf 101}, 203902 (2008).

\bibitem{Longhi09LPR}
S. Longhi, Laser and Photon. Rev. {\bf 3}, 243 (2009).

\bibitem{K1}
P.L. Kapitza, Sov. Phys. JETP {\bf 21}, 588 (1951).

\bibitem{K2}
L.D. Landau and E.M. Lifshitz, {\it Mechanics} (Pergamon, Oxford,
1960), pp. 93-95.

\bibitem{K3}
W. Paul, Rev. Mov. Phys. {\bf 62}, 531 (1990).

\bibitem{K4}
R.J. Cook, D.G. Shankland, and A.L. Wells, Phys. Rev. A {\bf 31},
564 (1985).

\bibitem{K5}
I. Gilary, N. Moiseyev, S. Rahav, and S. Fishman, J. Phys.  A {\bf
36}, L409 (2003).

\bibitem{Fan09}
L. Verslegers, P.B. Catrysse, Z. Yu, and S. Fan, Phys. Rev. Lett.
{\bf 103}, 033902 (2009).

\bibitem{Longhi07}
S. Longhi, Phys. Rev. B {\bf 76}, 195119 (2007).

\bibitem{note}
For a parabolic potential, Eqs.(3) reduce to the pendulum equations;
the two fixed points correspond to the unstable and stable
stationary points of the pendulum.

\bibitem{Lederer02}
T. Pertsch, T. Zentgraf, U. Peschel, A. Br\"{a}uer, and F. Lederer,
Phys. Rev. Lett. {\bf 88}, 093901 (2002).




%
%
%
%
%
%
%
%
%
%
%
%
%
%
%
%
%
%
%
%
%
%
%

\end{thebibliography}
\end{document}